\documentclass[camera,letterpaper,nomarginnotes,nonarrowgutter]{jpaper}
\usepackage[sort,compress]{cite}
\usepackage{amsmath,amssymb,amsfonts}
\usepackage{graphicx}
\usepackage{textcomp}
\usepackage{xcolor}
\usepackage[table]{xcolor}
\usepackage{xspace}
\usepackage{fancyhdr}
\usepackage{balance}
\usepackage{titling}
\usepackage[bookmarks=true,breaklinks=true,letterpaper=true,colorlinks,citecolor=blue,linkcolor=blue,urlcolor=blue]{hyperref}
\usepackage{algorithmic}
\usepackage{booktabs}
\usepackage{multirow}
\usepackage{comment}
\usepackage{float}
\usepackage{makecell}

\def\BibTeX{{\rm B\kern-.05em{\sc i\kern-.025em b}\kern-.08em
    T\kern-.1667em\lower.7ex\hbox{E}\kern-.125emX}}

\newcommand{\name}{Sudoku\xspace}
\newcommand{\eg}{\emph{e.g.}\xspace}
\newcommand{\ie}{\emph{i.e.}\xspace}

\newcommand{\subtitle}[1]{\posttitle{\par\normalfont{#1}\par\end{center}}}

\title{\Large{Sudoku: Decomposing DRAM Address Mapping into Component Functions}}

\author{
Minbok~Wi$\dagger$ \quad
Seungmin~Baek$\dagger$ \quad
Seonyong~Park$\dagger$ \quad
Mattan~Erez$\ddagger$ \quad
Jung~Ho~Ahn$\dagger$
\vspace{-2mm}
\\\\\emph{Seoul National University}$\dagger$ \qquad
\emph{The University of Texas at Austin}$\ddagger$
}

\begin{document}
\bstctlcite{IEEEexample:BSTcontrol}
\maketitle


\thispagestyle{plain}
\pagestyle{plain}




\begin{abstract}
Decomposing DRAM address mappings into component-level functions is critical for understanding memory behavior and enabling precise RowHammer attacks, yet existing reverse-engineering methods fall short.
We introduce novel timing-based techniques leveraging DRAM refresh intervals and consecutive access latencies to infer component-specific functions.
Based on this, we present \name, the first software-based tool to automatically decompose full DRAM address mappings into channel, rank, bank group, and bank functions while identifying row and column bits.
We validate \name's effectiveness, successfully decomposing mappings on recent Intel and AMD processors.
\end{abstract}

\section{Introduction}
\label{sec:introduction}

DRAM address mapping is a key feature of memory controllers (MCs), which translates memory requests into physical locations in DRAM to maximize parallelism and minimize contention in the memory system.
Simultaneously, the growing threat of DRAM-related attacks increases the importance of precise and detailed knowledge of the DRAM address mapping~\cite{2016-sec-drama, 2023-dramaqueen-dramsec, 2020-sp-trrespass, 2022-sp-blacksmith, 2024-sec-zenhammer, 2024-sec-sledgehammer, 2025-asplos-marionette}.
However, despite its importance for both performance and security, DRAM address mappings remain undocumented by processor manufacturers, necessitating efficient reverse-engineering methods.

Prior reverse-engineering methods~\cite{2016-sec-drama, 2020-dac-dramdig, 2023-dramsec-amdre, 2024-sec-zenhammer} primarily use the well-known row-buffer conflict timing channel to recover bank addressing functions; however, they require physical probing or fail to fully identify all row and column bits.
Furthermore, prior methods are unable to decompose DRAM address mapping into component functions,\footnote{We denote the memory system elements---such as the channel, DIMM, and rank---and the logical elements of a DRAM chip---such as the bank group and bank---collectively as \emph{components}. We use ``components'' and ``internal components'' interchangeably.} which becomes more important considering recent RowHammer attacks that exploit specific DRAM internal components~\cite{2023-sosp-siloz, 2024-sec-sledgehammer, 2024-isca-dramscope, 2025-asplos-marionette}.

In this paper, we revisit timing channels in DRAM-based memory systems to enable component-level decomposition of DRAM address mapping. 
First, we use DRAM refresh intervals as an indicator for inferring the granularity of refresh operations, known as refresh groups. 
We can determine whether two addresses are mapped to the same refresh group by alternating two memory addresses, detecting refresh-induced latency spikes, and measuring their intervals. 
Second, we analyze the latency of consecutive memory accesses and utilize this to infer the functions of the DRAM component. 
Memory access patterns affect the latency of consecutive memory accesses, depending on the target DRAM components, and provide information for identifying bank group and address functions.
Lastly, we briefly cover the non-uniqueness of the derived system of the DRAM addressing functions.

Based on our exploration, we develop \name, a software-based tool that identifies row and column bits and decomposes DRAM address mappings into component functions, even without physical access to systems.
\name successfully finds row and column bits and decomposes DRAM address mapping into component functions across recent Intel and AMD processors under various memory configurations.
We also verify our results based on the processor's MC-related registers and recent physical probing-based results of DRAM address mappings~\cite{2025-sec-mcsee, 2024-sec-zenhammer}.
Lastly, we open-source our code to aid future research on reverse-engineering the DRAM address mappings.\footnote{\href{https://github.com/scale-snu/sudoku.git}{https://github.com/scale-snu/sudoku.git}}

This paper makes the following key contributions:
\begin{itemize}
    \item We analyze timing channels in modern DRAM-based memory systems based on refresh intervals and consecutive memory accesses.
    \item We develop \name, a software-based tool that discovers full DRAM address mappings and decomposes these mappings into DRAM internal component functions, based on our timing channel analyses.
    \item We identify row and column bits without requiring physical access to the system, and we verify the results by confirming the injectivity of the derived mapping.
    \item We find DRAM component functions in recent Intel and AMD processors with various memory configurations.
\end{itemize}
\section{Background}
\label{sec:background}

\subsection{DRAM Organization, Operations, and Timings}
\label{subsec:background_dram}

\begin{figure}[!tb]
    \centering
    \includegraphics[width=0.99\columnwidth]{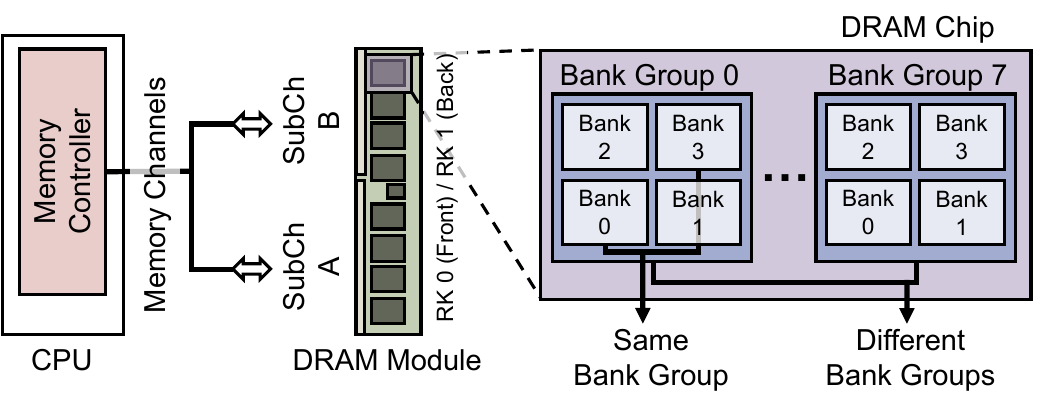}
    \caption[DRAM-based main memory systems]{DRAM-based memory system organization. DDR5 introduces sub-channels and doubles the number of bank groups compared to DDR4.}
    \label{fig:background_memory_system}
\end{figure}

DRAM-based main memory systems employ a hierarchical structure---typically comprising multiple \textit{components}, such as channels, DIMMs, ranks, bank groups, and banks---to achieve high parallelism (see Figure~\ref{fig:background_memory_system}).
Modern standards such as DDR5~\cite{jedec-ddr5} further increase complexity by introducing sub-channels and doubling the number of bank groups compared to DDR4~\cite{jedec-ddr4}.
Accessing data requires a Memory Controller (MC) to issue commands following specific timing constraints.
An \texttt{ACT} (activate) command opens a target row within a bank.
After a delay (\texttt{tRCD}), \texttt{RD} (read) or \texttt{WR} (write) commands access specific columns in the open, activated row.
Accessing a different row within the same bank necessitates a \texttt{PRE} (precharge) command and incurs an additional delay (\texttt{tRP}) prior to the next activation.
This sequence results in higher latency for same-bank accesses compared to row hits, creating the well-known row-buffer conflict timing channel.

Beyond row-buffer conflicts, other DRAM operations create observable timing variations~\cite{2020-sp-trrespass}.
DRAM cells require periodic refresh operations (\texttt{REF}) to retain data, which temporarily block memory accesses for \texttt{tRFC} cycles and occur at an average interval of \texttt{tREFI}.
Modern DRAM devices support various types of refresh operations, such as all-bank, fine-grained, or same-bank refreshes, to reduce performance overhead from refresh operations~\cite{jedec-ddr4, jedec-ddr5}.
MCs may implement those various refresh schemes, potentially affecting access latency differently depending on the target addresses.

Crucially for our work, DRAM employs mandatory timing parameters for consecutive accesses, such as \texttt{tRRD} (row-to-row delay) and \texttt{tCCD} (column-to-column delay), which vary depending on whether the accesses target the same or different bank groups~\cite{jedec-ddr4, jedec-ddr5}.
However, rather than directly using these DRAM timing parameters, MCs use their own set of timing parameters---such as \texttt{tRDRD}, \texttt{tRDWR}, \texttt{tWRRD}, and \texttt{tWRWR}---based on the sequence and type of memory accesses~\cite{intel-12th-datasheet, intel-13th-datasheet, amd-bkdg}.
These timing parameters are more complex, as they depend not only on the bank group and bank, but also on the rank and DIMM.
Moreover, they are non-SPD related timings whose values vary based on the system platform and configuration, and are typically determined during DDR training for optimal performance.
While often small, these timing differences between consecutive accesses represent another potential channel for inferring address mapping details.

\subsection{XOR-Based Hash Functions}
\label{subsec:background_xor}

Modern MCs typically use XOR-based hash functions to translate physical addresses into DRAM component indices (\eg, channel, rank, bank group, bank, row, and column)~\cite{2005-tc-xorhash, 2024-ccs-principled, 2024-ccs-coloring}.
Each XOR-based hash function generates a single-bit hash value by XORing a selected subset of physical address bits.
Thus, each function can be represented as a bitmask, where each set bit indicates that the corresponding address bit is involved in the hash function~\cite{2005-tc-xorhash}.
The complete mapping can be represented as a system of linear equations over $GF(2)$, often implemented as a binary matrix, designed to distribute memory accesses evenly~\cite{2005-tc-xorhash, 2024-ccs-coloring, 2024-ccs-principled}.

Reverse-engineering these undocumented functions relies on observing system behavior.
Prior methods employ brute-force~\cite{2016-sec-drama, 2023-dramsec-amdre, 2024-sec-zenhammer} or educated-guessing~\cite{2020-dac-dramdig, 2024-ccs-coloring, 2024-ccs-principled} techniques, primarily analyzing row-buffer conflicts to deduce relationships between physical address bits and bank/row mapping.
They generate random address pairs, observe conflicts (or lack thereof), and attempt to solve the underlying linear system with given input-output pairs.

While useful for identifying bank and some row/column bits, these conflict-based approaches generally cannot decompose the mapping into functions corresponding to memory systems' architectural components (\eg, channel, rank, and bank group).
Determining this full component-level decomposition is essential for precisely modeling memory behavior and enabling advanced security analyses, motivating the exploration of additional timing channels in this paper.

\section{Timing Channels for Component Function Identification}
\label{sec:identifying}

In this section, we analyze DRAM timing channels beyond simple row conflicts, focusing on refresh operations and consecutive accesses to infer component-level mapping details.

\subsection{Understanding How Systems Configure Memory via System Registers}
\label{subsec:identifying_understanding}

Accurate analysis requires understanding system-specific memory configurations and timings.
We identify these by examining BIOS settings and reading processor-specific MC-related registers~\cite{intel-12th-datasheet, intel-13th-datasheet, amd-bkdg}.
For example, we obtain precise values for key timing parameters such as \texttt{tRFC}, \texttt{tREFI}, and various \texttt{tRDRD} timings, which are essential for our subsequent analyses.
We also examine the MC-related registers that are no longer documented in recent processors but were disclosed in the previous datasheets, as in~\cite{2022-sp-msrtemplating}, to obtain several configurations related to DRAM address mapping~\cite{intel-12th-datasheet, intel-13th-datasheet}.

Knowing the system-configured refresh interval (\texttt{tREFI}) is particularly important, as it depends on the type of refresh operations and DRAM chip density~\cite{jedec-ddr4, jedec-ddr5}.
We also note that on the tested Intel processors with DDR5 (Table~\ref{tbl:tested_system}), the MC treats DDR5 sub-channels as two independent memory channels; a single-rank DDR5 module is perceived as two channels, and populating two physical channels results in the MC managing four logical channels.
This understanding informs our timing analyses on systems with DDR5.

Modern processors also provide limited information about DRAM address mapping through  MC-related registers~\cite{intel-12th-datasheet, intel-13th-datasheet}, enabling verification of specific DRAM component mapping functions.
For example, Intel Core processors reveal a channel hash mask and the bit used for bank group selection through their MC-related registers~\cite{intel-12th-datasheet, intel-13th-datasheet}.
Also, in Linux, the EDAC (Error Detection and Correction) subsystem utilizes these MC-related registers to accurately diagnose the DRAM error locations, thereby revealing portions of DRAM address mappings in the server processors~\cite{linux-edac, 2023-sosp-siloz}.
We leverage this DRAM address mapping information to validate \name's results (Section~\ref{sec:results}).

Finally, while accessing system registers aids our timing channel analysis, our \name tool does not require it.

\begin{figure}[!tb]
    \centering
    \includegraphics[width=0.99\columnwidth]{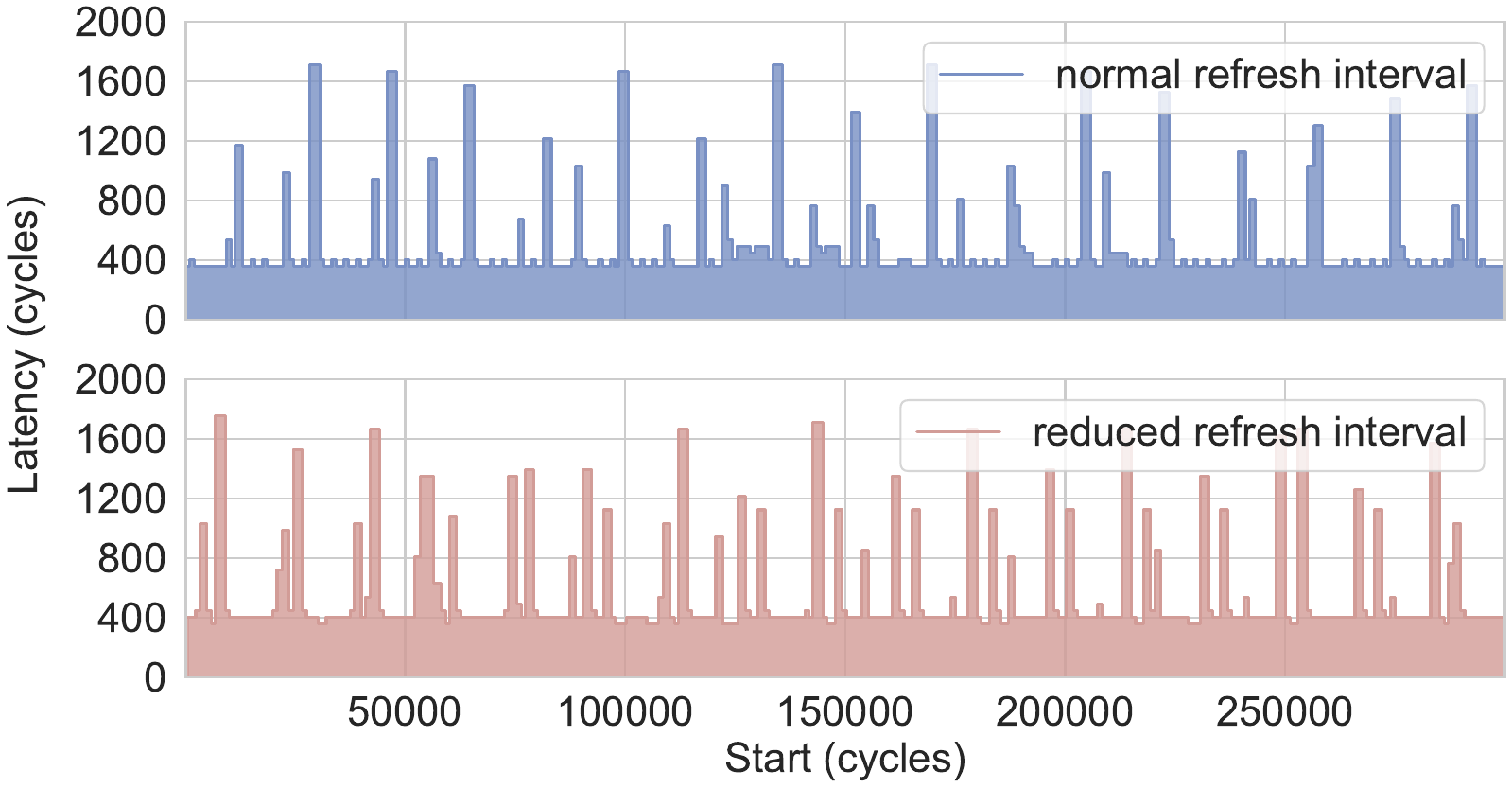}
    \caption[auto-refresh]{
        Measured refresh-induced latency spikes in the AMD Ryzen 9 7950X processor with DDR5 operating at 4.5GHz. 
        We iteratively access two memory addresses, measure their latencies, and compute the average spike interval.
        Two types of refresh intervals are shown: a normal refresh interval (top) and a reduced refresh interval (bottom).
    }
    \label{fig:identifying_refreshes}
\end{figure}

\subsection{Refresh Interval}
\label{subsec:identifying_refresh}

Beyond simple detection, we exploit refresh-induced latency spikes to infer refresh group---whether two addresses share the same refresh resources.
By alternately accessing a pair of randomly generated memory addresses and monitoring the interval between periodic latency spikes~\cite{2020-sp-trrespass, 2024-sec-zenhammer}, we observe distinct patterns.
Addresses mapped to the same refresh group yield spikes at the standard \texttt{tREFI} interval.
By contrast, addresses mapped to different refresh groups produce two interleaved spike trains, resulting in a reduced effective interval between observed latency spikes.
Figure~\ref{fig:identifying_refreshes} shows measured memory access latencies over time in the AMD Ryzen 9 7950X processor with DDR5.
For the normal refresh interval, we observe monolithic and periodic latency spikes. 
In contrast, for the reduced refresh interval, there are two distinct periodic spikes with a certain offset.
This directly reveals whether the address pair differs in the address bits determining the refresh group.

Applying this method, we observed all-bank refresh on an Intel 12th Gen processor with DDR4, fine-grained refresh on Intel 12th (Alder Lake) and 14th (Raptor Lake Refresh) Gen processors with DDR5, and all-bank refresh operations on an AMD Ryzen Zen 4 processor with DDR5.\footnote{Measurements on AMD systems required careful thresholding due to limited granularity of measuring timings and various noise factors.}
Furthermore, since XOR-based hash functions tend to partition the address space evenly, the proportion of random address pairs exhibiting reduced intervals provides an estimate of the number of address bits (and thus, XOR functions) involved in determining the refresh group.

\subsection{Consecutive Memory Accesses}
\label{subsec:identifying_consecutive}

\begin{figure}[!tb]
    \centering
    \includegraphics[width=0.99\columnwidth]{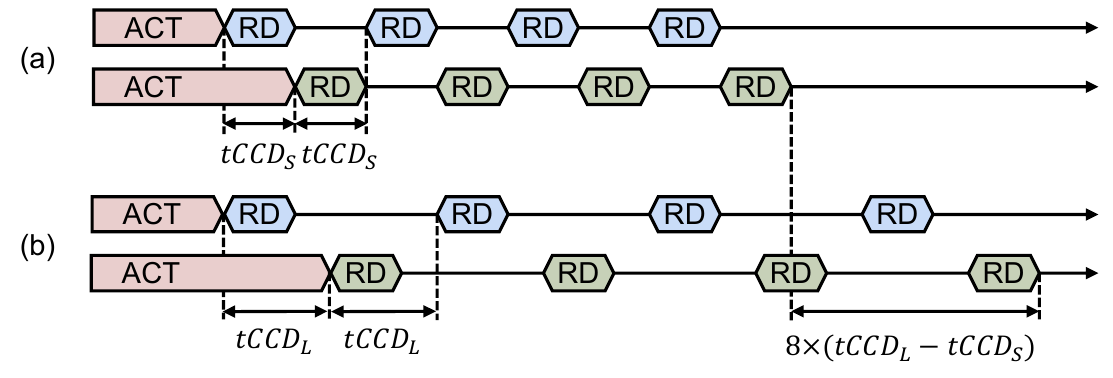}
    \caption[Consecutive Memory Accesses]{
        Theoretical timing differences between two consecutive memory reads. 
        Consecutive reads exhibit different timing characteristics depending on their mappings. 
        (a) Reads to different bank groups ($tCCD_{S}$) incur a shorter interval than (b) reads to the same bank group ($tCCD_{L}$).}
    \label{fig:identifying_consecutive_memory_accesses}
\end{figure}

MC's memory timing parameters, such as the variants of \texttt{tRDRD}, specify different minimum intervals for consecutive memory accesses depending on whether they target the same bank group, different bank groups, different ranks, or different DIMMs in both Intel and AMD processors~\cite{intel-12th-datasheet, intel-13th-datasheet, amd-bkdg}.
While these timing differences encode component mapping information, they are typically small (a few nanoseconds) and easily obscured by MC request scheduling and system noise.

To overcome this limitation, we amplify these subtle differences.
Our method creates and alternately accesses two \emph{memory streams}, each carefully constructed to ensure internal row-buffer hits.
This interleaving magnifies the base timing differences dictated by the relevant \texttt{tRDRD} variant between the two streams, as illustrated conceptually in Figure~\ref{fig:identifying_consecutive_memory_accesses}.

\begin{figure}[!tb]
    \centering
    \includegraphics[width=0.99\columnwidth]{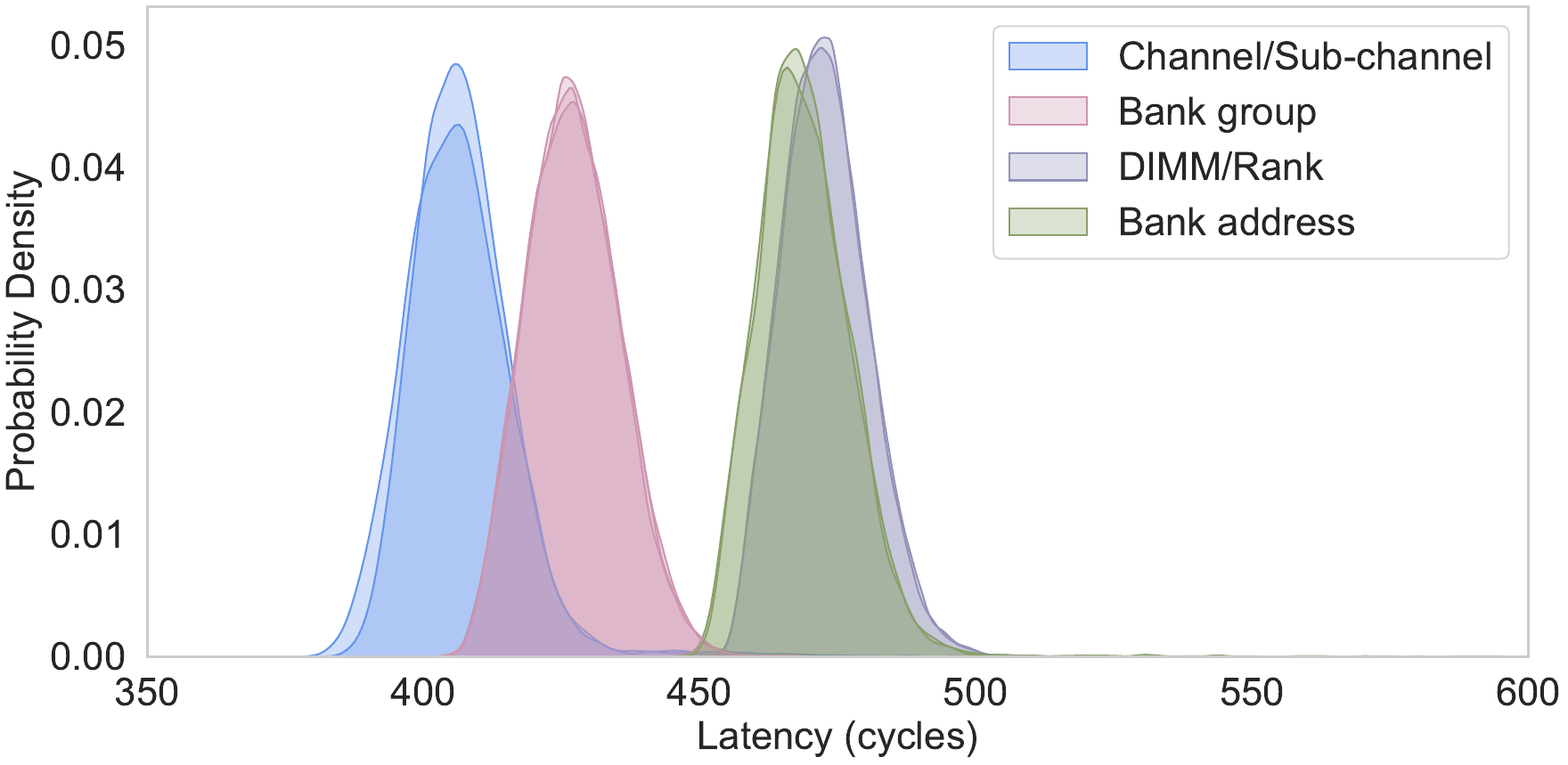}
    \caption[consecutive memory accesses]{
        The distribution of consecutive memory access latencies across memory address pairs that differ in only one of the nine DRAM addressing functions.
    }
    \label{fig:name_consecutive}
\end{figure}

Measuring the latency of these alternating stream accesses reveals distinct latency distributions (shown for an Intel Core i9-14900K in Figure~\ref{fig:name_consecutive}).
The peaks of these distributions correspond to the timing differences associated with accessing different component levels.
By comparing the measured peak latencies to the system-configured \texttt{tRDRD} values (obtained as per Section~\ref{subsec:identifying_understanding}), we can group the underlying XOR functions based on the component they likely map to (\eg, channel, sub-channel, bank group, DIMM/rank, and bank address).
Combined, refresh interval analysis and consecutive access timing allow us to group address mapping functions by component type.
However, the current method has limitations; system-configured MC timing parameters highly affect the latency of consecutive memory accesses.
For example, distinguishing rank and module functions could be ambiguous if their respective \texttt{tRDRD} timings are configured identically by the system.

\section{Sudoku}
\label{sec:sudoku}

\begin{figure}[!tb]
    \centering
    \includegraphics[width=0.99\columnwidth]{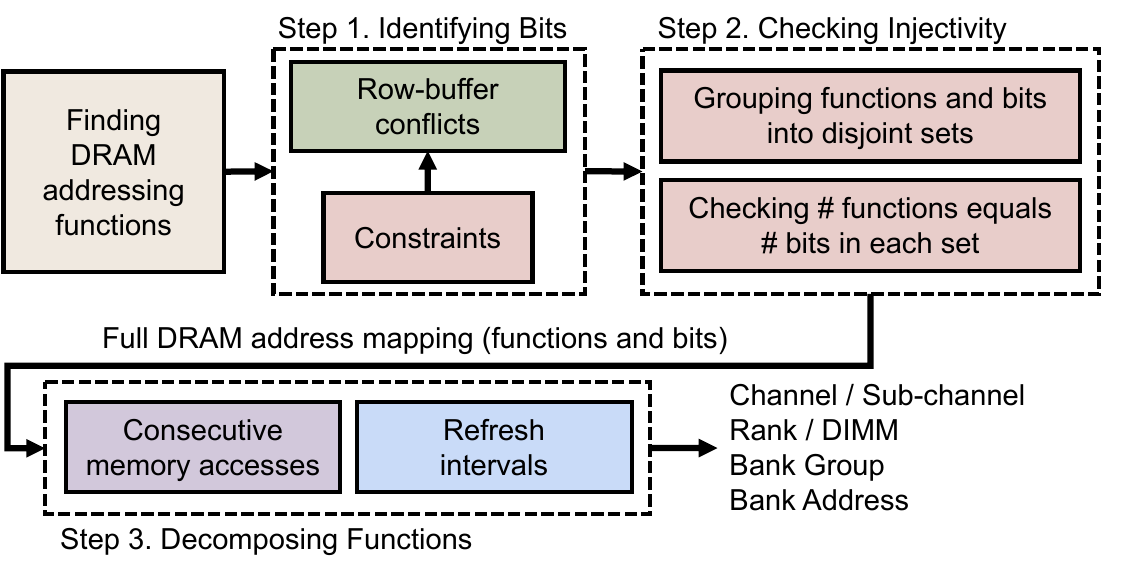}
    \caption[sudoku overview]{
        An overview of \name. 
        \name i) leverages row-buffer conflicts with constraints to identify row and column bits, ii) validates the DRAM address mapping by checking injectivity, and iii) exploits refresh intervals and consecutive memory accesses to decompose DRAM address mapping.}
    \label{fig:name_overview}
\end{figure}

We develop \name,\footnote{The tool is named \name because decomposing DRAM addressing functions resembles solving a \name puzzle, where each element must adhere to specific constraints.} a software tool for reverse-engineering DRAM address mapping in commercial computer systems, requiring no physical access.
\name extends existing tools (\eg, DRAMA~\cite{2016-sec-drama} and ZenHammer~\cite{2024-sec-zenhammer}) that reverse-engineer DRAM addressing functions using row-buffer conflicts.
Specifically, \name takes the DRAM addressing functions produced by these tools, automatically identifies row and column bits, and decomposes DRAM address mappings into component functions.
As depicted in Figure~\ref{fig:name_overview}, \name integrates conflict-based testing~\cite{2016-sec-drama} with timing channel analyses from Section~\ref{sec:identifying}.

\subsection{Generating Desired Memory Addresses}
\label{subsec:sudoku_address}

To effectively isolate the timing effects of specific DRAM components (for decomposition) or row/column bits (for identification), \name requires carefully crafted memory access patterns.
Instead of random testing or brute-forcing, \name solves the (partially) known mapping system to generate address pairs that differ only in the output of specific target functions, while keeping other function outputs identical.
This targeted approach is crucial for correctly attributing observed timing variations.

\name \emph{identifies row and column bits} using row-buffer conflicts as an oracle, guided by an educated-guessing approach similar to prior work to reduce the search space~\cite{2020-dac-dramdig, 2024-ccs-coloring}.
The key constraint is generating address pairs mapping to the same bank (\ie, having identical outputs for all bank-related hash functions).
\name systematically tests the role of each physical address bit by generating bitmasks.
It considers masks derived from the known functions and combinations involving previously unused address bits to ensure comprehensive coverage.
By generating address pairs based on these masks (XORing the mask with a base address) and observing whether accesses cause row-buffer conflicts or hits, \name classifies the corresponding physical address bits in the mask as contributing to row or column indexing, respectively. 

\subsection{Validating the System of Hash Functions}
\label{subsec:sudoku_validating}

\begin{figure}[!tb]
    \centering
    \includegraphics[width=0.99\columnwidth]{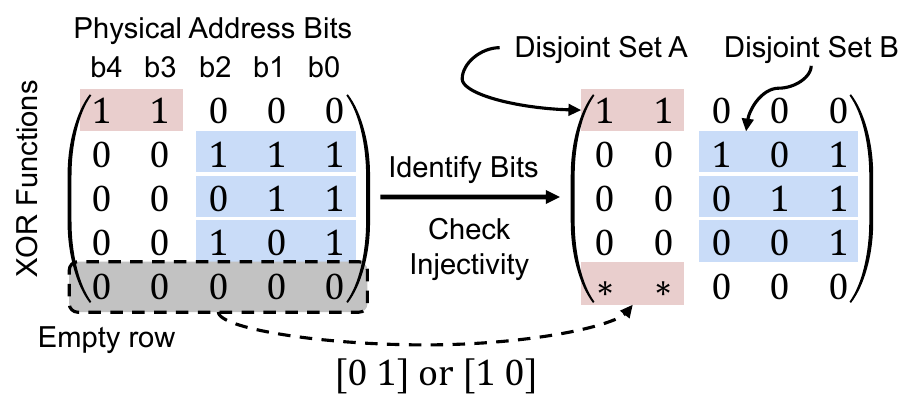}
    \caption[non-uniqueness]{The process of verifying injectivity and finding an additional function. Disjoint set A (subsystem) doesn't satisfy the injectivity, meaning that an additional function or bit is required.
    }
    \label{fig:name_injectivity}
\end{figure}

A valid DRAM address mapping must be \emph{injective} (one-to-one) to ensure correctness.
\name verifies this property by representing the derived functions and involved bits as a linear system over $GF(2)$ and applying the rank-nullity theorem~\cite{2005-tc-xorhash}.
\name decomposes the system into disjoint subsystems and checks if, for each subsystem, the number of linearly independent functions (rank) equals the number of involved physical address bits.

However, this conflict-based testing has inherent limitations due to the non-uniqueness of XOR-based hash systems~\cite{2005-tc-xorhash, 2024-ccs-coloring} (as illustrated in Figure~\ref{fig:name_injectivity}).
It may not be possible to uniquely determine every function if multiple function/bit combinations produce the same conflict behavior.
To resolve such ambiguities, \name follows common assumptions (\eg, assigning high-order bits to rows and low-order bits to columns~\cite{2016-sec-drama, 2020-dac-dramdig, 2024-sec-zenhammer}), ensuring the final mapping preserves injectivity and functional correctness, even if alternative valid representations might exist~\cite{2005-tc-xorhash, 2024-ccs-coloring}.
Lastly, we avoid excessive function reduction (\eg, reducing the system to a minimal basis) that could alter the original conflict characteristics relevant for analysis when considering both DRAM address mapping functions and row/column bits.

\subsection{Decomposing DRAM Address Mapping}
\label{subsec:name_decomposing}

In its final step, \name decomposes the validated set of DRAM addressing functions, assigning each function to its corresponding physical component
(\eg, channel, rank, bank group, and bank address).
This decomposition leverages the distinct timing signatures identified in Section~\ref{sec:identifying}, measured using the targeted memory access patterns generated as described in Section~\ref{subsec:sudoku_address}.
\name systematically tests each relevant XOR function (or group of functions potentially related to a component) using the following timing analyses: refresh intervals and consecutive memory accesses.

\noindent
\textbf{Refresh Intervals:}
\name generates pairs of addresses that differ only in the output bits produced by the target function(s).
By measuring the interval between refresh-induced latency spikes when accessing these pairs (as detailed in Section~\ref{subsec:sudoku_address}), \name identifies the functions that control the refresh group.
Functions causing a ``reduced'' refresh interval pattern are marked as related to the refresh scope defined by the system configuration (\eg, channel/sub-channel, rank/DIMM or potentially finer granularities, depending on the refresh type employed by the system).

\begin{table}[tb!]
    \centering
    \caption[Tested system configurations]{Tested system configurations.}
    \label{tbl:tested_system}
    \resizebox{1\columnwidth}{!}{
        \begin{tabular}{cccc}
            \toprule
                \textbf{System} & \textbf{Processor (microcode)} & \textbf{Motherboard} & \textbf{Memory Devices}\\
            \toprule
                \multirow{2}{*}{Intel-A} &
                    Intel Core & ASUS & 32 GB DDR4-3200 \\
                    & i9-12900K (\texttt{0x38}) & Z690-A & 2R$\times$8 UDIMM \\
                \cmidrule(l){2-4}
                \multirow{2}{*}{Intel-B} &
                    Intel Core & ASUS & 32 GB DDR5-4800 \\
                    & i9-12900K (\texttt{0x38}) & Z690-F & 2R$\times$8 UDIMM \\
                \cmidrule(l){2-4}
                \multirow{2}{*}{Intel-C} &
                    Intel Core & MSI & 32 GB DDR5-4800 \\
                    & i9-14900K (\texttt{0x12C}) & B760M & 2R$\times$8 UDIMM \\
            \midrule
                \multirow{2}{*}{AMD-A} &
                    AMD Ryzen 9 & ASRock & 32 GB DDR5-4800 \\  
                    & 7950X (\texttt{0xA601206}) & X670E & 2R$\times$8 UDIMM \\
            \bottomrule
        \end{tabular}
    }
    \vspace{0.15in}
\end{table}
\begin{table*}[tb!]
    \centering
    \caption{Reverse-engineered DRAM addressing functions and decomposed results using \name. 
    N/A denotes that there is no function for the corresponding component.
    Each value in the table represents a mask for an XOR-based hash function, where the parity of the set bits in the mask determines the output.}
    \label{tbl:sudoku_results}
\resizebox{\textwidth}{!}{
    \begin{tabular}{ccllllllll}
    \toprule
    \multirow{4}{*}{\textbf{System}} & 
        \multirow{4}{*}{
            \begin{tabular}[c]{@{}c@{}}
                \textbf{Memory}\\ 
                \textbf{Configuration}
            \end{tabular}
        } &
        \multicolumn{6}{c}{
            \textbf{Functions}
        } & 
        \multicolumn{2}{c}{
            \textbf{Bits}
        } \\
    \cmidrule(lr){3-8} \cmidrule(lr){9-10}
    & & 
        \multicolumn{1}{c}{
            \multirow{3}{*}{
                \begin{tabular}[c]{@{}c@{}}
                    \textbf{Channel and}\\ 
                    \textbf{Sub-Channel}
                \end{tabular}
            }
        } & 
        \multicolumn{1}{c}{
            \multirow{3}{*}{
                \begin{tabular}[c]{@{}c@{}}
                    \textbf{DIMM and}\\ 
                    \textbf{Rank}
                \end{tabular}
            }
        } & 
        \multicolumn{4}{c}{\textbf{Banks}} & 
        \multicolumn{1}{c}{
            \multirow{3}{*}{\textbf{Row}}
        } & 
        \multicolumn{1}{c}{
            \multirow{3}{*}{\textbf{Column}}
        } \\
    \cmidrule(lr){5-8}
    & & & & 
        \multicolumn{2}{c}{\textbf{Bank Group}} & 
        \multicolumn{2}{c}{\textbf{Bank Address}} & & \\
    \cmidrule(lr){1-1} \cmidrule(lr){2-2} \cmidrule(lr){3-3} \cmidrule(lr){4-4} \cmidrule(lr){5-6} \cmidrule(lr){7-8} \cmidrule(lr){9-9} \cmidrule(lr){10-10}
    \multirow{6}{*}{\textbf{Intel-A}} & 
        1Ch-1DPC & 
        \texttt{N/A} & 
        \texttt{0x0000088000} & 
        \texttt{0x0000002A00}, &
        \texttt{0x0124044000} & 
        \texttt{0x0249910000}, &
        \texttt{0x0492620000} & 
        \texttt{0x07FFFC0000} & 
        \texttt{0x0000001FC0} \\
    & 
        \cellcolor{gray!15}1Ch-2DPC & 
        \cellcolor{gray!15}\texttt{N/A} & 
        \cellcolor{gray!15}\texttt{0x0000108000}, &
        \cellcolor{gray!15}\texttt{0x0000002A00}, & 
        \cellcolor{gray!15}\texttt{0x0924084000} & 
        \cellcolor{gray!15}\texttt{0x0249210000}, & 
        \cellcolor{gray!15}\texttt{0x0492840000} & 
        \cellcolor{gray!15}\texttt{\texttt{0x0FFFF80000}} & 
        \cellcolor{gray!15}\texttt{0x0000001FC0} \\
    & 
        \cellcolor{gray!15} & 
        \cellcolor{gray!15} & 
        \cellcolor{gray!15}\texttt{0x0000420000} & 
        \cellcolor{gray!15} & 
        \cellcolor{gray!15} & 
        \cellcolor{gray!15} & 
        \cellcolor{gray!15} & 
        \cellcolor{gray!15} & 
        \cellcolor{gray!15} \\
    & 
        2Ch-1DPC & 
        \texttt{0x0000082600} & 
        \texttt{0x0000110000} & 
        \texttt{0x0000005400}, &
        \texttt{0x0248088000} & 
        \texttt{0x0493220000}, &
        \texttt{0x0924C40000} & 
        \texttt{0x0FFFF80000} & 
        \texttt{0x0000001FC0} \\
    & 
        \cellcolor{gray!15}2Ch-2DPC & 
        \cellcolor{gray!15}\texttt{0x0000082600} & 
        \cellcolor{gray!15}\texttt{0x0000210000}, &
        \cellcolor{gray!15}\texttt{0x0000005400}, &
        \cellcolor{gray!15}\texttt{0x1248108000} & 
        \cellcolor{gray!15}\texttt{0x0492420000}, &
        \cellcolor{gray!15}\texttt{0x0925080000} & 
        \cellcolor{gray!15}\texttt{0x1FFFF00000} & 
        \cellcolor{gray!15}\texttt{0x0000001FC0} \\
    &
        \cellcolor{gray!15} &
        \cellcolor{gray!15} & 
        \cellcolor{gray!15}\texttt{0x0000840000} &
        \cellcolor{gray!15} & 
        \cellcolor{gray!15} & 
        \cellcolor{gray!15} & 
        \cellcolor{gray!15} & 
        \cellcolor{gray!15} & 
        \cellcolor{gray!15} \\
    \cmidrule(l){1-10}
    \multirow{8}{*}{
        \begin{tabular}[c]{@{}c@{}}
            \textbf{Intel-B,}\\ 
            \textbf{Intel-C}
        \end{tabular}} & 
        1Ch-1DPC & 
        \texttt{0x00000C3200} & 
        \texttt{0x0000410000} & 
        \texttt{0x0000081100}, &
        \texttt{0x0222104000}, & 
        \texttt{0x0088820000}, &
        \texttt{0x0111040000} & 
        \texttt{0x07FFF80000} & 
        \texttt{0x0000000FC0} \\
    &
        &
        &
        &
        &
        \texttt{0x0442080000} &
        &
        &
        &
        \\
    &
        \cellcolor{gray!15}1Ch-2DPC & 
        \cellcolor{gray!15}\texttt{0x00000C3200} & 
        \cellcolor{gray!15}\texttt{0x0000810000}, &
        \cellcolor{gray!15}\texttt{0x0000081100}, & 
        \cellcolor{gray!15}\texttt{0x0222104000}, &
        \cellcolor{gray!15}\texttt{0x0114100000}, &
        \cellcolor{gray!15}\texttt{0x088A020000} & 
        \cellcolor{gray!15}\texttt{0x0FFFE80000} & 
        \cellcolor{gray!15}\texttt{0x0000000FC0} \\
    &
        \cellcolor{gray!15} &
        \cellcolor{gray!15} &
        \cellcolor{gray!15}\texttt{0x0001040000} &
        \cellcolor{gray!15} &
        \cellcolor{gray!15}\texttt{0x0444408000} & 
        \cellcolor{gray!15} &
        \cellcolor{gray!15} &
        \cellcolor{gray!15} &
        \cellcolor{gray!15} \\
    & 
        2Ch-1DPC & 
        \texttt{0x0000104200}, & 
        \texttt{0x0000820000} & 
        \texttt{0x0000102100}, &
        \texttt{0x0444208000}, &
        \texttt{0x0111040000}, &
        \texttt{0x0222080000} & 
        \texttt{0x0FFFF00000} & 
        \texttt{0x0000001BC0} \\
    &
        &
        \texttt{0x0000186400} &
        &
        &
        \texttt{0x0888410000} &
        &
        &
        &
        \\
    &
        \cellcolor{gray!15}2Ch-2DPC & 
        \cellcolor{gray!15}\texttt{0x0000104200}, &
        \cellcolor{gray!15}\texttt{0x0001020000}, &
        \cellcolor{gray!15}\texttt{0x0000102100}, &
        \cellcolor{gray!15}\texttt{0x0444408000}, &
        \cellcolor{gray!15}\texttt{0x0228200000}, &
        \cellcolor{gray!15}\texttt{0x1114040000} &
        \cellcolor{gray!15}\texttt{0x1FFFD00000} &
        \cellcolor{gray!15}\texttt{0x0000001BC0} \\
    &
        \cellcolor{gray!15} &
        \cellcolor{gray!15}\texttt{0x0000186400} &
        \cellcolor{gray!15}\texttt{0x0002080000} &
        \cellcolor{gray!15} &
        \cellcolor{gray!15}\texttt{0x0888810000} &
        \cellcolor{gray!15} &
        \cellcolor{gray!15} &
        \cellcolor{gray!15} &
        \cellcolor{gray!15}\\
    \midrule
    \multirow{8}{*}{\textbf{AMD-A}} & 
        1Ch-1DPC &
            \texttt{0x07FFF80040} &
            \texttt{0x0000040000} &
            \texttt{0x0084200100}, &
            \texttt{0x0108400200}, &
            \texttt{0x0042100800}, &
            \texttt{0x0421080400} &
            \texttt{0x07FFF80000} &
            \texttt{0x000003E080} \\
        &
            &
            &
            &
            &
            \texttt{0x0210801000} &
            &
            &
            &
            \\
        &
            \cellcolor{gray!15} 1Ch-2DPC &
            \cellcolor{gray!15}\texttt{0x0FFFF00040} &
            \cellcolor{gray!15}\texttt{0x0000040000}, &
            \cellcolor{gray!15}\texttt{0x0108400100}, &
            \cellcolor{gray!15}\texttt{0x0210800200}, &
            \cellcolor{gray!15}\texttt{0x0084200800}, &
            \cellcolor{gray!15}\texttt{0x0842100400} &
            \cellcolor{gray!15}\texttt{0x0FFFF00000} &
            \cellcolor{gray!15}\texttt{0x000003E080} \\
        &
            \cellcolor{gray!15} &
            \cellcolor{gray!15} &
            \cellcolor{gray!15}\texttt{0x0000080000} &
            \cellcolor{gray!15} &
            \cellcolor{gray!15}\texttt{0x0421001000}&
            \cellcolor{gray!15} &
            \cellcolor{gray!15} &
            \cellcolor{gray!15} &
            \cellcolor{gray!15} \\
        & 
            2Ch-1DPC &
            \texttt{0x0000000100}, &
            \texttt{0x0000080000} &
            \texttt{0x0108400200}, &
            \texttt{0x0210800400}, &
            \texttt{0x0084201000}, &
            \texttt{0x0842100800} &
            \texttt{0x0FFFF00000} &
            \texttt{0x000007C080} \\
        &
            &
            \texttt{0x0FFFF00040} &
            &
            &
            \texttt{0x0421002000}&
            &
            &
            &
            \\
        &
            \cellcolor{gray!15}2Ch-2DPC &    
            \cellcolor{gray!15}\texttt{0x0000000100}, &
            \cellcolor{gray!15}\texttt{0x0000080000}, &
            \cellcolor{gray!15}\texttt{0x0210800200}, &
            \cellcolor{gray!15}\texttt{0x0421000400}, &
            \cellcolor{gray!15}\texttt{0x0108401000}, &
            \cellcolor{gray!15}\texttt{0x1084200800} &
            \cellcolor{gray!15}\texttt{0x1FFFE00000} &
            \cellcolor{gray!15}\texttt{0x000007C080} \\
        &
            \cellcolor{gray!15} &
            \cellcolor{gray!15}\texttt{0x1FFFE00040} &
            \cellcolor{gray!15}\texttt{0x0000100000} &
            \cellcolor{gray!15} &
            \cellcolor{gray!15}\texttt{0x0842002000} &
            \cellcolor{gray!15} &
            \cellcolor{gray!15} &
            \cellcolor{gray!15} &
            \cellcolor{gray!15} \\
    \bottomrule
    \end{tabular}
}
\end{table*}

\noindent
\textbf{Consecutive Memory Accesses:}
To further differentiate functions, especially those not distinguished by refresh or those necessitating finer classification (like bank group vs. bank address),
\name employs the consecutive access timing analysis (Section~\ref{subsec:identifying_consecutive}).
It generates alternating memory streams where the underlying addresses differ only based on the target function(s).
The resulting latency distribution is analyzed; the location of the primary latency peak is compared against the known \texttt{tRDRD} timing parameters of the system.
A match allows \name to associate the target function(s) with the corresponding component level (\eg, mapping a function to bank group if its activation leads to latency that is consistent with \texttt{tRDRD} timing for different bank groups).

By systematically applying these two timing analyses to the reverse-engineered functions using precisely controlled address generation, \name labels each function with its inferred component role.
This process yields the decomposed DRAM address mapping, detailing how different sets of physical address bits map to channel, sub-channel, rank/DIMM, bank group, and bank address indices.

The accuracy of this decomposition depends on the system's behavior and configuration, as noted previously.
Factors such as the MC's refresh strategy, request scheduling policies, and whether timing parameters for different component interactions are distinct and measurable can \emph{limit} the ability to differentiate certain functions 
(\eg, channel vs.\ sub-channel, or rank vs.\ DIMM).

\section{Results}
\label{sec:results}

\noindent
\textbf{Experimental setup:} 
We evaluated \name on the recent Intel and AMD systems detailed in Table~\ref{tbl:tested_system}.
We used memory access latencies in CPU cycles, measured using the \texttt{rdtscp} instruction with core dynamic frequency scaling disabled to ensure stable measurements~\cite{2016-sec-drama, 2024-sec-zenhammer}.
For channel/sub-channel functions, which are often directly specified or easily inferred from MC-related registers (MCHBAR), we used these known values directly to constrain the reverse-engineering process, without affecting the overall mapping functionality.
Lastly, we regarded recent physical probing-based DRAM address mapping results~\cite{2024-sec-zenhammer, 2025-sec-mcsee} as ground truth and verify that \name generates consistent DRAM address mapping results.
Full results, including the derived XOR masks, are presented in Table~\ref{tbl:sudoku_results}.

\noindent
\textbf{Intel Core Processors:} 
\name successfully decomposed the DRAM address mapping for the tested Intel Core processors with the given DRAM addressing functions derived using prior tool~\cite{2016-sec-drama}.
Consistent with trends observed in AMD processors~\cite{2024-sec-zenhammer} and differing from some older Intel architectures~\cite{2016-sec-drama, 2020-dac-dramdig}, these recent Intel platforms utilize most identified physical address bits within their mapping functions.
Notably, for DDR5 configurations, we observed discontinuities in the physical address bits used for row and column indexing, influenced by multi-channel and multi-DPC (DIMM per channel) setups.
Our timing analyses also confirmed the use of fine-grained all-bank refresh with DDR5, contrasting with the standard all-bank refresh observed with DDR4 on the same platform, demonstrating \name's ability to reveal such configuration details.
Lastly, it is worth noting that \name can identify full DRAM address mapping, thus requiring no physical access to the system.

\noindent
\textbf{AMD Zen 4 Processor:}
On the AMD Zen 4 platform, considering the necessary PCI address offset as described in prior work, \name produced decomposed mappings consistent with those reported by ZenHammer~\cite{2024-sec-zenhammer}.
The tool successfully distinguished rank/DIMM and sub-channel functions using refresh intervals and bank group and bank functions via consecutive memory accesses.
The AMD processors issue separate refresh commands for each sub-channel, each DIMM, and each rank, which means that it is possible to distinguish between channel and sub-channel functions.

\section{Conclusion}
\label{sec:conclusion}

This paper has demonstrated novel techniques for decomposing undocumented DRAM address mappings by exploiting previously underutilized timing channels.
We first showed that DRAM refresh intervals reveal refresh group, and second, that amplified timing variations in consecutive memory accesses expose component-level mapping information.
Leveraging these insights, we developed \name, a software-based tool that automatically decomposes DRAM addressing functions into their specific component roles (\eg, channel, rank, bank group, and bank).
We validated its effectiveness on recent Intel and AMD processors.
This work provides crucial, fine-grained mapping information necessary for advanced RowHammer analysis and vulnerability assessment on modern platforms. 
By open-sourcing Sudoku, we hope to enable further research into DRAM security and system behavior.

\section*{Acknowledgment}

We thank Hwayong Nam and Michael Jaemin Kim at Seoul National University for their valuable feedback to the paper.

\bibliographystyle{IEEEtranS}
\bibliography{ref}

\end{document}